\documentclass[aps,pra,a4paper,10pt,twocolumn,superscriptaddress]{revtex4-1} 
\usepackage{appendix,graphicx,subfigure,amsmath,amssymb,hyperref,mathrsfs,amsmath,amsfonts,amsthm,amssymb}
\usepackage{subfigure}

\usepackage{color} 
\usepackage{xcolor} 
\usepackage{soul,ulem,cancel} 
\usepackage{epstopdf} 

\newcommand{\op}[1]{\hat{#1}}

\newcommand{\qmean}[1]{\langle{#1}\rangle}
\begin{document}

\title{Optomechanical oscillator controlled by variation in its heat bath temperature}

\author{Michal Kol\'a\v r}
\affiliation{Palack\'{y} University, Department of Optics, 17.~listopadu 1192/12, 771~46~Olomouc, Czech Republic}
\author{Artem Ryabov}
\affiliation{
Charles University, Faculty of Mathematics and Physics, Department of Macromolecular Physics, V Hole{\v s}ovi{\v c}k{\' a}ch 2, 180~00~Praha~8, Czech Republic}
\author{Radim Filip}
\affiliation{Palack\'{y} University, Department of Optics, 17.~listopadu 1192/12, 771~46~Olomouc, Czech Republic}
\date{\today }

\begin{abstract}
We propose a generation of a low-noise state of optomechanical oscillator by a temperature dependent force. We analyze the situation in which a quantum optomechanical oscillator (denoted as the membrane) is driven by an external force (produced by the piston). Both systems are embedded in a common heat bath at certain  temperature $T$. The driving force the piston exerts on the membrane is bath temperature dependent. Initially, for $T=T_0$, the piston is linearly coupled to the membrane. The bath temperature is then reversibly changed to $T\neq T_0$. The change of temperature shifts the membrane, but simultaneously also increases its fluctuations.
The resulting equilibrium state of the membrane is analyzed from the point of view of mechanical, as well as of thermodynamic characteristics. We compare these characteristics of membrane and derive their intimate connection. Next, we cool down the thermal noise of the membrane, bringing it out of equilibrium, still being in the contact with heat bath. This cooling retains the effective canonical Gibbs state with the effective temperature $T^\star$. In such case we study the analogs of the equilibrium quantities for low-noise mechanical states of the membrane.
\end{abstract}

\maketitle
\section{Introduction}
\label{intro}
A preparation of low-noise and coherent quantum states of a broad class of macroscopic mechanical systems is currently a bottleneck of quantum opto-electro-mechanics \cite{aspelmeyer}. A preparation of quantum mechanical
coherent states of mirrors, membranes and nanostructures by coherent optical and microwave driving is available \cite{verhagen}. Recently, the motion of levitating nanospheres is approaching quantum mechanical ground state \cite{jain}. To prepare a mechanical state with large
coherent amplitude and low thermal noise, the current methods use simultaneous cooling of the mechanical oscillator and large intensity coherent states of light or microwave field. Coherent state of light or microwave radiation simply drive mechanical system to a coherent mechanical state \cite{glauber}. Optically driven nonlinear mechanisms can also generate self-oscillations interpreted as mechanical lasing \cite{grudinin}. All these processes are very different from the ones in a shot-noise limited laser, where thermal fluctuations of environment generate a high-quality coherent state due to a nonlinearity of saturable lasing mechanism \cite{scully}. Thermal classical noise, pumping the laser, can be converted to low-noise coherent states of light with a variance of electric field fluctuations invariant over a phase for an arbitrary amplitude. It is an autonomous process without any use of other coherent source that qualifies laser as a primary source for many
applications in metrology, nonlinear optics, quantum optics and quantum communication. It can be stimulating for a development of a mechanical system autonomously generating low-noise coherent state from thermal fluctuations of an environment, rather than from optical or
microwave coherent pump. In this case, light can be used only to
independently measure characteristics of thermally excited coherent quantum mechanical oscillator.

Thermal expansion can be an obvious candidate to provide a small displacement of tiny mechanical system by heating \cite{taylor}. Thermal expansion is a tendency of a solid, liquid or gas to change its dimensions in response to a change in temperature and it is characterized by a coefficient of thermal
expansion. Heat can also produce mechanical changes indirectly, for example, through Seebeck thermoelectric effect \cite{disalvo},
magneto-Seebeck effect \cite{walter}, spin-Seebeck effect \cite{uchida}, or photo-thermal effect \cite{cerdonio}. For majority of matter, liquid or gas systems, the length $L$ at $T_0$ increases about $\Delta L$ by temperature difference $\Delta T$, typically linearly $\Delta L/L=\alpha \Delta T$ to a first approximation. 
\begin{figure}[htb]
\centering \hspace{-0.06\linewidth}
\includegraphics[width=.9\linewidth]{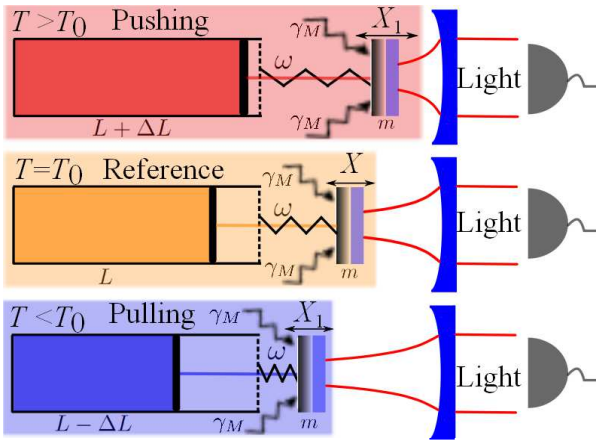}
\caption{Schematic of the analyzed opto-thermo-mechanical
setup. The optomechanical membrane and the thermomechanical piston are both embedded in a heat bath with the coupling strength $\gamma_M$ at temperature $T$. At some reference temperature $T_0$ the thermomechanical piston has a length $L$, setting some reference equilibrium position $\qmean{X}=0$ of the membrane (with mass $m$ and angular frequency $\omega$). By changing the bath temperature $T$ we shift the equilibrium position of the membrane to $\qmean{X_1}=0$ due to the piston thermal expansion. The light field serves merely for readout of the membrane state and potentially, to cool the membrane down as is discussed in the Appendix. 
} \label{figure-gedanken}
\end{figure}
The process is overdamped, loosing quickly thermal energy to
a reservoir. However, the system is simultaneously heated by higher
temperature $T_0+\Delta T$ of the reservoir, which increases a Brownian
noise in the process. The noise with a small amplitude is then displacing
position of tiny microscopic high-quality mechanical oscillator, membrane
or nanoobject. In this ideal limit, any back action of the oscillator on
the process can be neglected. A larger temperature difference $\Delta T$
means simultaneously also the larger variance of Brownian noise in the
process. This reduces the quality of the oscillator. It is therefore principally important to find the optimal overall
performance of such thermomechanical process which is prospective for an
autonomous steady-state generation of mechanical coherent states. 

We analyze the situation in which an optomechanical  oscillator, our system under study, is linearly coupled to another external system, the thermomechanical piston. We use this term motivated by a gedanken experiment with a gas chamber and piston depicted in Fig.~\ref{figure-gedanken}. In general, it can be any temperature-dependent actuator based on different physical principle. The piston drives the membrane by a constant force, parametrically dependent on the heat bath temperature $T$ surrounding both the membrane and the piston. By changing the heat bath temperature the driving changes the average position of the membrane, as well as fluctuations are added due to the membrane coupling to the heat bath. We use light only to  independently monitor statistical changes of mechanical motion and confirm our mechanical and thermodynamic prediction. The read out of mechanical position and momentum can be done by standard methods of quantum optomechanics \cite{aspelmeyer-book}.

We have found that the process described above can be  capable (for realistic values of the relevant parameters) to prepare a low-noise, displaced thermal state just by reversibly manipulating a {\it single macroscopic} parameter, namely the bath temperature $T$. By the low-noise state we denote a state with larger average position squared than its position variance caused by a change of some external parameter. Such behavior is quantified by the Signal-to-Noise Ratio (SNR) of the position. The only externally controlled parameter that is changed in our model is the heat bath temperature $T$. It is known that one can drive the membrane by means of other deterministic parameter different from temperature $T$, in order to reach high SNR, but characterization of the $T$-dependent driving  is missing. We show that the relevant variables describing our model will behave {\it qualitatively} differently in the case of $T$-dependent driving as compared to the case of $T$-independent parametric driving.
The change of the temperature $T$ brings the membrane into a new thermal equilibrium state. We study the properties of this state from the mechanical point of view as well as from the thermodynamic point of view. The temperature dependent displacement of the average position changes the potential energy of the membrane linked to the first moment of the position, hence, has a direct connection to the {\it work} done on the membrane. The second moment of the position is changed as well, being connected to the {\it heat} supplied to the membrane. We show a direct connection between the work and the mechanical SNR of the position. This connection naturally arises, as well, in the thermodynamic analysis. The temperature dependent SNR modifies some of the thermodynamic potentials describing the membrane by additional $T$-dependent terms. Thus, the $T$ derivatives of these quantities are modified with respect to the case of membrane driven by temperature independent constant force.  We discuss the impact on the membrane {\it heat capacity} $c(T)$ and the dependence of the obtained results on the used measurement method. Our results are based on probably the simplest (linear) model of situation described by the temperature dependent Hamiltonian. 

The paper is structured in the following way. Section~\eqref{model} describes the used model of the membrane and the transformation of the membrane state the piston does by means of the heat bath temperature $T$ change. 
In Section~\eqref{mechanics} we analyze the connection between the statistical moments of the membrane position and their contribution to the different forms of the potential energy, as well as the connection of SNR and the average total energy of the membrane. Section~\eqref{thermodynamics} describes the membrane by thermodynamic potentials and the relations between them in the presence of the temperature dependent drive by the piston. Section~\eqref{conclusions} concludes the results of the analysis and suggests further research.

\section{The model}
\label{model}
We model our system, denoted as the optomechanical membrane, as a quantum harmonic oscillator with the Hamiltonian $\op{H}_0$,
\begin{equation}
\op{H}_0=\frac{\op{P}^2}{2m}+\frac{m\omega^2}{2}\op{X}^2.
\label{eq-equilibrium-plain-oscill-ham0}
\end{equation} 
The membrane is  linearly driven by an external thermomechanical system, denoted as the piston, through the interaction Hamiltonian $\op{H}_I(T)$,
\begin{equation} 
\op{H}_I(T)=- f(T)\op{X},
\label{eq-equilibrium-plain-oscill-hamI}
\end{equation}
where  $f(T_0)=0$, for some fixed reference temperature $T_0$. 
{\color{red} }
The total Hamiltonian determining the dynamics of the membrane, $\op{H}(T)=\op{H}_0+\op{H}_I(T)$, thus reads
\begin{equation}
\label{eq-equilibrium-ham}
\op{H}(T)=\frac{\op{P}^2}{2m}+\frac{m\omega^2}{2}\left[\op{X}-\frac{f(T)}{m\omega^2}\right]^2-\frac{f(T)^2}{2m\omega^2},
\end{equation}
where we explicitly assume throughout the paper that the membrane frequency $\omega$ is kept {\it constant}.
Furthermore we assume that the membrane is weakly coupled to a heat bath at temperature $T$, which thermalizes the membrane into the canonical state with respect to the eigenbasis of $\op{H}(T)$. 

\begin{figure}[htb]
\centering \hspace{-0.06\linewidth}
\includegraphics[width=.9\linewidth]{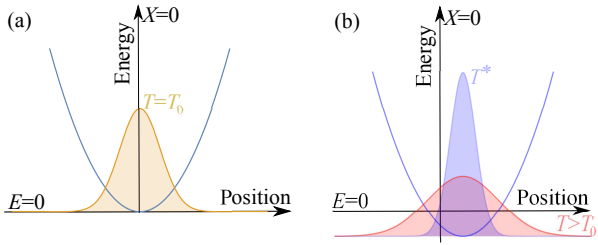}
\caption{Sketch of the model assumed in the main text.  (a) The harmonic oscillator  (membrane) in contact with a heat bath (temperature $T_0$, orange) is described by the Hamiltonian $\op{H}_0$, Eq.~\eqref{eq-equilibrium-plain-oscill-ham0}, initially. After switching on the temperature dependent interaction, $\op{H}_I(T_0)=- f(T_0)\op{X}$ with the external piston, at this temperature $T_0$, (b) we change the temperature of the heat bath to $T>T_0$, as  indicated (red).  The quadratic potential shifts horizontally (new equilibrium position), as well as vertically (new ground state energy), cf. Eq.~\eqref{eq-equilibrium-ham}. Alternativelly, we can switch on an external cooling ( see the main text) to bring the membrane to the state with an effective temperature $T^\star$ (blue), Eq.~\eqref{eq-eff-temp-definition}. } \label{figure-model}
\end{figure}

We will analyze the following transformation. We switch on the temperature dependent interaction, $\op{H}_I(T_0)$, at some initial temperature $T_0$, keeping the position of the potential minimum and the zero energy level, see Fig.~\ref{figure-model}, unchanged with respect to the $\op{H}_0$, Eq.~\eqref{eq-equilibrium-plain-oscill-ham0}. 
The membrane with the bare Hamiltonian $\op{H_0}$ at $T=T_0$ is taken as a reference with respect to the position and value of the potential minimum. 

Subsequently, we change the bath temperature from $T_0$ to $T$, thus shifting the membrane position (the square bracket term in 
Eq.~\eqref{eq-equilibrium-ham}) and the energy offset (the last term in 
Eq.~\eqref{eq-equilibrium-ham}) with respect to the reference state. 
This change in temperature 
also regulates intensity of thermal noise acting on the membrane. By heating, the membrane can be pushed further from the initial average position, increasing the stored elastic energy, but at the same time, the uncertainty of membrane position increases. This trade-off is the key effect, which will be analyzed from both the mechanical and thermodynamic point of view.
In another words, the  temperature dependence of $f(T)$ establishes the interconnection between the membrane coherent shift and absorbed heat due to the change of $T$.  Any external $T$-independent drive shifts the membrane as well, thus is capable of creating a low-noise (high SNR) mechanical state, but one needs at least one more independent parameter. We study the consequences of the assumption that change of $T$ shifts the membrane and changes its fluctuations jointly, thus being the only external parameter that is changed.

From a general perspective, the present scenario belongs to a broad class of models with effective temperature-dependent Hamiltonians (or energy levels)
\cite{rushbrooke}, \cite{shental}. In such models, first discussed in detail in Refs.~\cite{rushbrooke}, the temperature-dependence of effective Hamiltonians occur due to a partial averaging over possible microstates of a large equilibrium systems. Contrary to this, in our case, the temperature dependence explicitly enters $\op{H}(T)$ through the {\it external} driving force exerted by the expanding thermomechanical device. 
The model in Eq.~\eqref{eq-equilibrium-ham} can be regarded as a temperature dependent analogy of the situation encountered in \cite{steeneken}. 

\section{Mechanical characteristics}
\label{mechanics}
As outlined above, we are primarily interested in quantifying the trade-off between the elastic (or coherent) energy stored in the membrane as compared to the amount of noise it gains from the thermal bath. Both aspects are to be discussed in thermal equilibrium, which is the simplest situation that could be realized experimentally. 
The mechanical point of view allows for access (through an appropriate measurement) to the statistical moments of $X$ and $P$ of the membrane (hats are omitted in the following). 
 
The thermal equilibrium state is given by the Hamiltonian \eqref{eq-equilibrium-ham}, cf. Eq.~\eqref{eq-appA-gibbs}. According to the current experimental realizations \cite{aspelmeyer,steeneken} the high temperature limit, $k_BT\gg \hbar\omega$, is assumed in the rest of the paper. In this limit, the average values and variances of the membrane variables are
\begin{eqnarray}
\label{eq-equilibrium-average}\qmean{P}&=&0,\;\qmean{X}=\frac{f(T)}{m\omega^2},\\
{\rm Var}(X)&=&\frac{1}{m^2\omega^2}\qmean{P^2}=\frac{k_BT}{m\omega^2},
\label{eq-equilibrium-variance}
\end{eqnarray}
where $\langle \cdot \rangle$ represents the averaging over the equilibrium Gibbs state for a given temperature $T$ and given coupling $f$, Eq.~\eqref{eq-equilibrium-ham}.  Larger value of average position can be reached for smaller mass $m$ and smaller angular frequency $\omega$ of the membrane. Note that the variances in Eq.~\eqref{eq-equilibrium-variance} do {\it not } depend on the piston driving $f(T)$. They remain the same as for equilibrium mechanical oscillator without any external force. From the mechanical point of view, using Eq.~\eqref{eq-equilibrium-variance}, we can also define contributions of different origin to the average value of the potential energy part of the Hamiltonian \eqref{eq-equilibrium-ham}. Namely, the {\it coherent} part of the potential energy $E_p^{\rm coh}$ originating from the powers of $\qmean{X}$ and the {\it incoherent} part of the potential energy $E_p^{\rm inc}$ originating from the Var$(X)$ term
\begin{eqnarray}
E_p^{\rm coh}=-\frac{m\omega^2}{2}\qmean{X}^2,\;E_p^{\rm inc}=\frac{m\omega^2}{2}{\rm Var}(X).\label{eq-equilibrium-pot-energies}
\end{eqnarray}

This terminology comes from quantum limit of the coherent mechanical states \cite{glauber}. If the system in the ground state is shifted by deterministic force towards a coherent quantum state with minimal Var$(X)$ =Var$(P)$, only coherent part of energy increases.  
From the result of Eq.~\eqref{eq-equilibrium-variance} we see, as well, that in principle the coherent shift of the membrane position mean value (the signal) can change faster with the temperature $T$ compared to the linear $T$ dependence of  the variance (the noise) of $X$.  Their ratio can thus increase above unity when the temperature $T$ is changed. These facts are captured by the $T$-dependent quantity, the signal to noise ratio (SNR), defined for the {\it  equilibrium} membrane state. Taking Eqs.~\eqref{eq-equilibrium-average}-\eqref{eq-equilibrium-variance}, we obtain
\begin{eqnarray}
{\rm SNR}_X(T)=\frac{\qmean{X}^2}{{\rm Var}(X)}=\frac{f(T)^2}{k_B Tm\omega^2}.
\label{eq-equilibrium-snr}
\end{eqnarray}
We assume the linear temperature dependence
\begin{eqnarray}
f(T)=\kappa\alpha (T-T_0),
\label{eq-equilibrium-fT-model1}
\end{eqnarray}
where $\alpha>0$ is some temperature independent coefficient of thermal expansion \cite{steeneken}, and $\kappa$ characterizes the strength of the mutual membrane piston coupling.
The SNR can then be written as
\begin{eqnarray}
{\rm SNR}_X(T)=\frac{T_0\kappa^2\alpha^2}{k_Bm\omega^2}\frac{(\theta-1)^2}{\theta},\;\theta=\frac{T}{T_0}.\label{eq-equilibrium-snr-highT}
\end{eqnarray}

\begin{figure}[htb]
\centering \hspace{-0.06\linewidth}
\includegraphics[width=.7\linewidth]{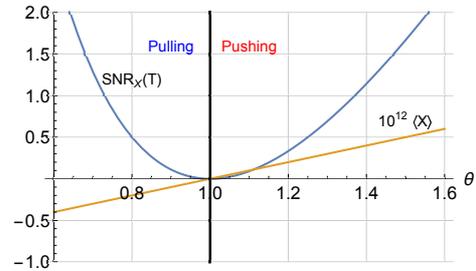}
\caption{The plot of the signal to noise ratio, Eq.~\eqref{eq-equilibrium-snr} and the rescaled average position $10^{12}\;\qmean{X}$, Eq.~\eqref{eq-equilibrium-average}, dependence on $\theta$, Eq.~\eqref{eq-equilibrium-snr-highT}. We used the model \eqref{eq-equilibrium-fT-model1} and the following values \cite{steeneken}, $\alpha \approx 10^{-12}\;{\rm m/K},\; m\omega^2\approx 10^{4}\; {\rm kg/s^2},\;\kappa/m\omega^2\approx 10^{-2},\;T_0\approx 10^{2}\;{\rm K}$. The non-monotonic and assymetric behavior shows that one has two options to increase the SNR$_X(T)$ above the desired value  SNR$_X(T)=1$. One can either increase the temperature ratio $\theta >1$ (pushing), thus increasing the  denominator in Eq.~\eqref{eq-equilibrium-snr} linearly as opposed to the quadratic increase  of the nominator. Even more advantageous option it to decrease the temperature ratio $\theta <1$ (pulling), in which the denominator decreases linearly, while the nominator again increases quadraticaly due to its symmetry .  } \label{figure-snr}
\end{figure}

The SNR$_X(T)$, Fig.~\ref{figure-snr}, describes how the quadratic displacement (the first moment) and the variance (the second moment) of the equilibrium state relatively change with the temperature. Apparently, larger $\mbox{SNR}_X (T)$ can be reached for smaller mass and smaller frequency, similarly to the mean position, Eq.~\eqref{eq-equilibrium-average}. The simple form of Eq.~\eqref{eq-equilibrium-snr} shows two options for increasing SNR$_X(T)$ above unity, meaning that the squared coherent shift is dominating over the variance increase. Thus, by manipulating the membrane surroundings temperature, while assuming the parametric $T$ dependence of the Hamiltonian \eqref{eq-equilibrium-ham}, we can increase the squared coherent shift  more than we add the thermal noise into the membrane. 
The first option, pushing, is to {\it increase} the temperature $T>T_0$, while increasing the average position squared $\qmean{X}^2$ {\it and} increasing the Var$(X)$, as well. The second option, pulling, is to {\it decrease} the temperature $T<T_0$, while {\it again} increasing the average position square $\qmean{X}^2$ {\it and} decreasing the Var$(X)$, Fig.~\ref{figure-snr}. 
An obvious asymmetry of SNR$_X(T)$ is due to the fact that the nominator of Eq.~\eqref{eq-equilibrium-snr-highT} increases symmetrically for $T\neq T_0$, i.e. the membrane shift is the same whether we increase or decrease the temperature, while the denominator is a monotonically increasing function of $T$, i.e., the higher the temperature the larger the variance of $X$. 
Both motions can be used to reach low-noise mechanical displacement, while pulling shows to be the preferable option. It is however at a cost of cooling the environment, which can be more demanding than heating it up.  
Note, that the assumption of the $T$-{\it independent} driving, $f\neq f(T)$, dictates SNR$_X(T)$, Eq.~\eqref{eq-equilibrium-snr},  being {\it inversely} proportional to $T$. Thus, for $T$-independent driving, one can also reach (low-noise) state with $\qmean{X}^2$ higher than Var($X$), increasing the SNR. But for the temperature {\it increase} the SNR will drop down in contrast to our $T$-dependent case.

When the piston drives the membrane, its position is changed with respect to the potential energy part of $H_0$, Eq.~\eqref{eq-equilibrium-plain-oscill-ham0}. This corresponds to the change in the membrane potential energy with respect to $H_0$, as well as change in the average value of the total Hamiltonian $\qmean{H(T)}$
\begin{eqnarray}
\qmean{H(T)} &=&\frac{\qmean{P^2}}{2m}+\frac{m\omega^2}{2}\qmean{X^2} - f(T)\qmean{X}.
\label{eq-equilibrium-average-E}
\end{eqnarray}
After some manipulations we obtain simple relation between the membrane total average energy and high $T$ limit \eqref{eq-equilibrium-snr} of SNR$_X(T)$ as
\begin{eqnarray}
\qmean{H(T)} =k_BT\left[ 1-\frac{{\rm SNR}_X(T)}{2}\right].
\label{eq-equilibrium-SNR-E-relation}
\end{eqnarray}
Equation~\eqref{eq-equilibrium-SNR-E-relation} shows that by {\it increasing} SNR$_X(T)\geq 0$ we always {\it decrease} the total average energy of the membrane. 
The second term in the square brackets of Eq.~\eqref{eq-equilibrium-SNR-E-relation} arises from the combination of the terms containing the first moment of the membrane position. This can be seen using the standard relation between the variance, the first, and the second statistical moment of $X$. The case in which SNR$_X(T)\geq 1$ shows that the square of the first moment $\qmean{X}$ has increased above the position variance Var$(X)$. Thus, the overall energy reduces below $k_B T/2$ and the potential energy becomes negative. 
From the point of view of average potential energy, it corresponds to the cancellation of the part of potential energy $E_p^{\rm coh}$ and $E_p^{\rm inc}$ in Eq.~\eqref{eq-equilibrium-pot-energies}. For $\mbox{SNR}_X(T)>2$, the total average energy becomes negative. It means, that the decrease of $E_p^{\rm coh}$ dominates over the increase of $E_p^{\rm inc}$. This negative potential and total energy appears however also for a linear, temperature independent force. We therefore focus on a change of potential energy  with the temperature. 

The contributions to potential energy, and the temperature derivative is 
\begin{eqnarray}
E_p^{\rm coh}(T)&=&-\frac{k_BT\,{\rm SNR}_X(T)}{2},\label{eq-equilibrium-coherent-potE}\\
\frac{{\rm d}E_p^{\rm coh}}{{\rm d}T}&=&-\frac{k_B}{2m\omega^2}\frac{{\rm d}f^2}{{\rm d}T},
\end{eqnarray}
underlining the fact that the coherent part of the potential energy can be manipulated through  the temperature change {\it only} if the membrane driving is temperature dependent. The first line of Eq.~\eqref{eq-equilibrium-coherent-potE} can be written as $E_p^{\rm coh}(T)-E_p^{\rm coh}(T_0)$, as well, see Eq.~\eqref{eq-equilibrium-fT-model1}. This allows for its interpretation as the {\it work W}, to be supplied to the membrane to change its average position from $\qmean{X}_{f=0}$ to $\qmean{X}$, cf. Eq.\eqref{eq-equilibrium-work}. The sign of the ${\rm d}E_p^{\rm coh}/{\rm d}T$ reflects the sign of the change of coherent part of the potential energy with $T$. If this derivative is zero, the function $f(T_0)$ can be zero for certain $T_0$ or ${\rm d}f/{\rm d}T$ is zero at $T_0$. This second possibility trivially includes the case $f\neq f(T)$, meaning that SNR$_X(T)$ decreases as $1/T$ for a fixed $f$. 

For comparison, the incoherent contribution to the membrane potential energy reads
\begin{eqnarray}
E_p^{\rm inc}(T)=\frac{k_BT}{2},
\label{eq-equilibrium-incoh-potE}
\end{eqnarray}
which again reflects the fact that the second moments of $X$ do not depend on the membrane driving $f(T)$, Eq.~\eqref{eq-equilibrium-variance}, hence one can not use functional dependencies of $f(T)$ to increase or decrease it. 

The connection between $\qmean{H(T)}$, SNR$_X(T)\geq 0$, shows direct dependence between the temperature dependent driving $f(T)$ and one of the fundamental equilibrium thermodynamic quantities, the internal energy of the system $U=\qmean{H(T)}$. Therefore, in the following we investigate the thermodynamic consequences of the temperature dependent driving of the membrane and how it affects the  relations between equilibrium thermodynamic potentials.
\section{Thermodynamics}
\label{thermodynamics}
Thermodynamic viewpoint of the driven membrane is different compared to the mechanical one, although the physical system is the same. Mechanical view on the physical system allows to measure position and momentum directly and evaluate their statistical moments. Therefore we can split energy to kinetic and potential part and further to coherent and incoherent parts. Thermodynamics does not distinguish these parts of energy.   
For example, from the energetic point of view, thermodynamics works with the total average energy $\qmean{H(T)}$, denoted as the internal energy $U$, for describing the equilibrium properties of the membrane state, thus it belongs to state variables. Other state variables of the membrane studied here are the free energy $F(T)$, and the entropy $S(T)$. 

\begin{figure}[htb]
\centering \hspace{-0.06\linewidth}
\includegraphics[width=.7\linewidth]{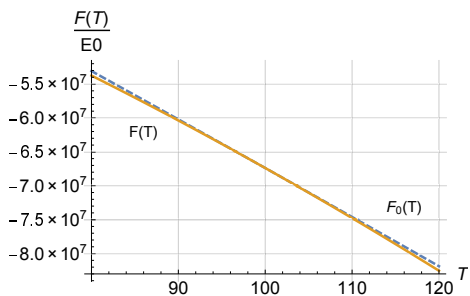}
\caption{The free energy $F$ and $F_0(T)$ (dashed), Eq.~\eqref{eq-equilibrium-potentials-F}, dependence on the thermodynamic temperature $T$ (in Kelvins). Energy unit $E_0$ is the energy of the membrane ground state. We used the model \eqref{eq-equilibrium-fT-model1} and the following values \cite{steeneken}, $\alpha \approx 10^{-12}\;{\rm m/K},\; m\omega^2\approx 10^{4}\; {\rm kg/s^2},\;\kappa/m\omega^2\approx 10^{-2},\;T_0\approx 10^{2}\;{\rm K}$.   } \label{figure-freeE}
\end{figure}

As in the previous section, we assume  the high temperature limit $k_BT\gg \hbar\omega$ in the following calculations.
For determination of the thermodynamic potentials we adopt the standard approach known from statistical mechanics \cite{greiner} and determine the partition function $Z(T)$ of the driven membrane as
\begin{eqnarray}
\label{eq-equlibrium-partition-f}
Z(T)&=&\exp\left[\frac{1}{k_BT}\frac{f(T)^2}{2m\omega^2}\right]Z_0(T)\\ \nonumber
&=&\exp\left[\frac{{\rm SNR}_X(T)}{2}\right]Z_0(T),\\
Z_0(T)&=&\sum_n \exp[-E_{0n}/k_BT],\\
Z(T)&=&\sum_n \exp[-E_n(T)/k_BT],
\end{eqnarray}
where $E_{0n}$ and $E_n(T)$ are the eigenvalues of the Hamiltonians $\op{H}_0$ and $\op{H}(T)$, Eqs.~\eqref{eq-equilibrium-plain-oscill-ham0}, \eqref{eq-equilibrium-ham}, respectively. The subscript ``$0$" stands in the following for standard textbook  \cite{greiner} situation referring to the thermodynamic quantities of a plain (undriven) harmonic oscillator. From Eq.~\eqref{eq-equlibrium-partition-f} we obtain the following  thermodynamic potentials including the corrections to the temperature dependent driving, see Figs.~\eqref{figure-freeE}-\eqref{figure-U}. The derivation for a more general Hamiltonian temperature-dependence is outlined in Appendix~\ref{sec-app-A}. Here we apply these results to the Hamiltonian \eqref{eq-equilibrium-ham}, and obtain
\begin{eqnarray}
U&=&U_0(T)-\frac{f(T)^2}{2m\omega^2}, \label{eq-equilibrium-potentials-U}\\
F&=&F_0(T)-\frac{f(T)^2}{2m\omega^2},\label{eq-equilibrium-potentials-F}\\
S&=&-\frac{\partial F_0}{\partial T}
,\label{eq-equilibrium-potentials-S}
\end{eqnarray}
where we have denoted $U_0(T)=\langle\op{H}_0\rangle_{f =0}$, $F_0(T)=-k_BT\ln Z_0$, 
and $\langle \cdot \rangle$ represents again the averaging over the equilibrium Gibbs state for a given temperature $T$ and the full Hamiltonian from Eq.~\eqref{eq-equilibrium-ham}. 

From Eqs.~\eqref{eq-equilibrium-potentials-U}-\eqref{eq-equilibrium-potentials-S} we see that the first two quantities have additional $T$-dependent terms stemming from the temperature dependent driving of the membrane by the piston. 
Namely, the internal energy $U=\qmean{H(T)}$ is modified due to the last term in Eq.~\eqref{eq-equilibrium-ham}, corresponding to the shift of the zero-energy level (see vertical shift of the parabolic potential in Fig.~\ref{figure-model}). If one is to measure the thermodynamic state quantities, or their equilibrium changes, e.g., the internal energy $U$ for systems defined by the $T$-dependent Hamiltonians, care should be taken of the method used. If the measurement would rely on the state (defined by the populations of the energy levels) of the membrane, we will obtain the result $U_0(T)$, or its appropriate change. On the other hand if the determination relies on the amount of energy transferred from the surroundings to the membrane, we will obtain as the result the change of $U$. Similarly it holds for the free energy $F$. On the contrary, the thermodynamic entropy $S$ of the membrane, Eq.~\eqref{eq-equilibrium-potentials-S}, remains unchanged with respect to the entropy of the {\it undriven} ($f=0$) membrane. 

The above mentioned possibility of obtaining two different results for the internal energy resulting from different methods of its measurement is directly related to multiple possibilities of getting another interesting thermodynamic characteristic of a system, namely the heat capacity. 
In the case of energy exchange with the surrounding heat bath, the heat $\delta Q_M$  enters the membrane and changes its internal energy by ${\rm d}U_0$. Such a method, e.g., the standard differential scanning calorimetry (DSC) yields as a result the heat capacity $c_0(T)$ given by
\begin{equation} 
c_0(T)=\frac{\delta Q_{{\rm M}}}{{\rm d} T}=\frac{{\rm d} U_0}{{\rm d} T} = T \frac{\partial S}{\partial T}.
\label{eq-equilibrium-heat-cap-entropy}
\end{equation}
Another possibility of obtaining $c_0(T)$ is to use light and measure statistics of position and momentum 
(as illustrated in Fig.~\ref{figure-gedanken}), which
goes beyond standard measurement of energy in thermodynamics. 


The second way how to define the capacity is by means of measurement of the total mean energy of the membrane, which is the mean value of the total Hamiltonian $\hat{H}(T)$, Eq.~\eqref{eq-equilibrium-ham}. Such capacity,  
\begin{eqnarray}
\label{eq-c-definition}
c(T)=\frac{{\rm d} U}{{\rm d} T}, 
\end{eqnarray}
describes change in total energy of the membrane when the heat bath temperature is varied. Of course, for temperature-independent Hamiltonians ($f\neq f(T)$), the both capacities $c(T)$ and $c_0(T)$ are equal. The difference between the two, i.e., the difference between $dU$ and $dU_0$, is caused by work performed on the membrane during heating, since $dU_0=\delta Q_M$, but $dU=\delta W + \delta Q_M$. Consequently, the capacity $c(T)$ can not be determined from the change of the membrane entropy, cf.\ last equality in Eqs.~\eqref{eq-equilibrium-heat-cap-entropy}. Instead, we have 
\begin{eqnarray}
c(T)= T\frac{\partial S}{\partial T} -\frac{1}{2m\omega^2}\frac{{\rm d}f^2}{{\rm d}T}.
\label{eq-equilibrium-heat-cap}
\end{eqnarray}
Whereas, standard assumption of quantum thermodynamics, on which we base our definition~\eqref{eq-c-definition} is that one can determine the mean energy of the system \cite{koslof}, it is an interesting open question how to actually measure the change in the mean energy (the capacity) for the thermally driven membrane without need to access the position and momentum statistics. 

\begin{figure}[htb]
\centering \hspace{-0.06\linewidth}
\includegraphics[width=.7\linewidth]{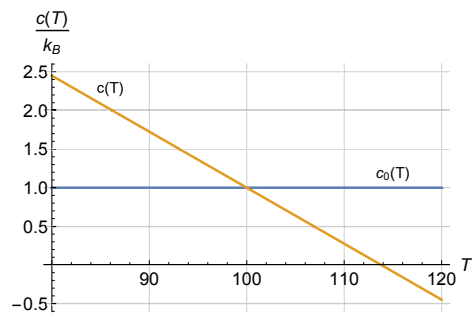}
\caption{The capacity $c (T)$, Eq.~\eqref{eq-equilibrium-heat-cap-model}, dependence on the thermodynamic temperature $T$ (in Kelvins). We used the following values \cite{steeneken}, $\alpha \approx 10^{-12}\;{\rm m/K},\; m\omega^2\approx 10^{4}\; {\rm kg/s^2},\;\kappa/m\omega^2\approx 10^{-2},\;T_0\approx 10^{2}\;{\rm K}$. The linear decrease of $c(T)$ reflects the extra energy that has to be supplied by the driving piston to the membrane while changing its potential, Eq.~\eqref{eq-equilibrium-ham}.  } \label{figure-heat-capacity}
\end{figure}

Equations \eqref{eq-equilibrium-heat-cap-entropy} and \eqref{eq-equilibrium-heat-cap} belong among main stimulating results of the paper. They show that systems with temperature dependent Hamiltonians allow for positive, zero or negative capacity.
As an example we can assume again $f(T)=\kappa\alpha (T-T_0)$,
 as in Eq.~\eqref{eq-equilibrium-snr-highT}, then
\begin{eqnarray}
c(T)&=&T\frac{\partial S}{\partial T} -\frac{\kappa^2 \alpha^2(T-T_0)}{m\omega^2},
\label{eq-equilibrium-heat-cap-model}
\end{eqnarray}
where the second term {\it changes} the sign when crossing the temperature $T_0$, see Fig.~\ref{figure-heat-capacity}. We see that the heat capacity \eqref{eq-equilibrium-heat-cap} of the temperature driven membrane is modified with respect to the undriven one.
This is a direct consequence of the fact that the internal energy $U$, Eq.~\eqref{eq-equilibrium-potentials-U}, has additional $T$-dependent term compared to the case of the bare harmonic oscillator with the Hamiltonian $H_0$, Eq.~\eqref{eq-equilibrium-plain-oscill-ham0}. 
For our particular choice of $f(T)$, the capacity decreases linearly with the temperature. It reflects the already mentioned interplay between the shift of the potential minimum and the increase of thermal fluctuations in the shifted parabolic potential. After the internal energy of the system passes through its maximum (where the decrease of the potential minimum exactly compensates the increase in thermal fluctuations), the capacity becomes negative since the average energy of our system decreases with increasing $T$. 

\begin{figure}[htb]
\centering \hspace{-0.06\linewidth}
\subfigure[]{
\includegraphics[width=0.48\linewidth]{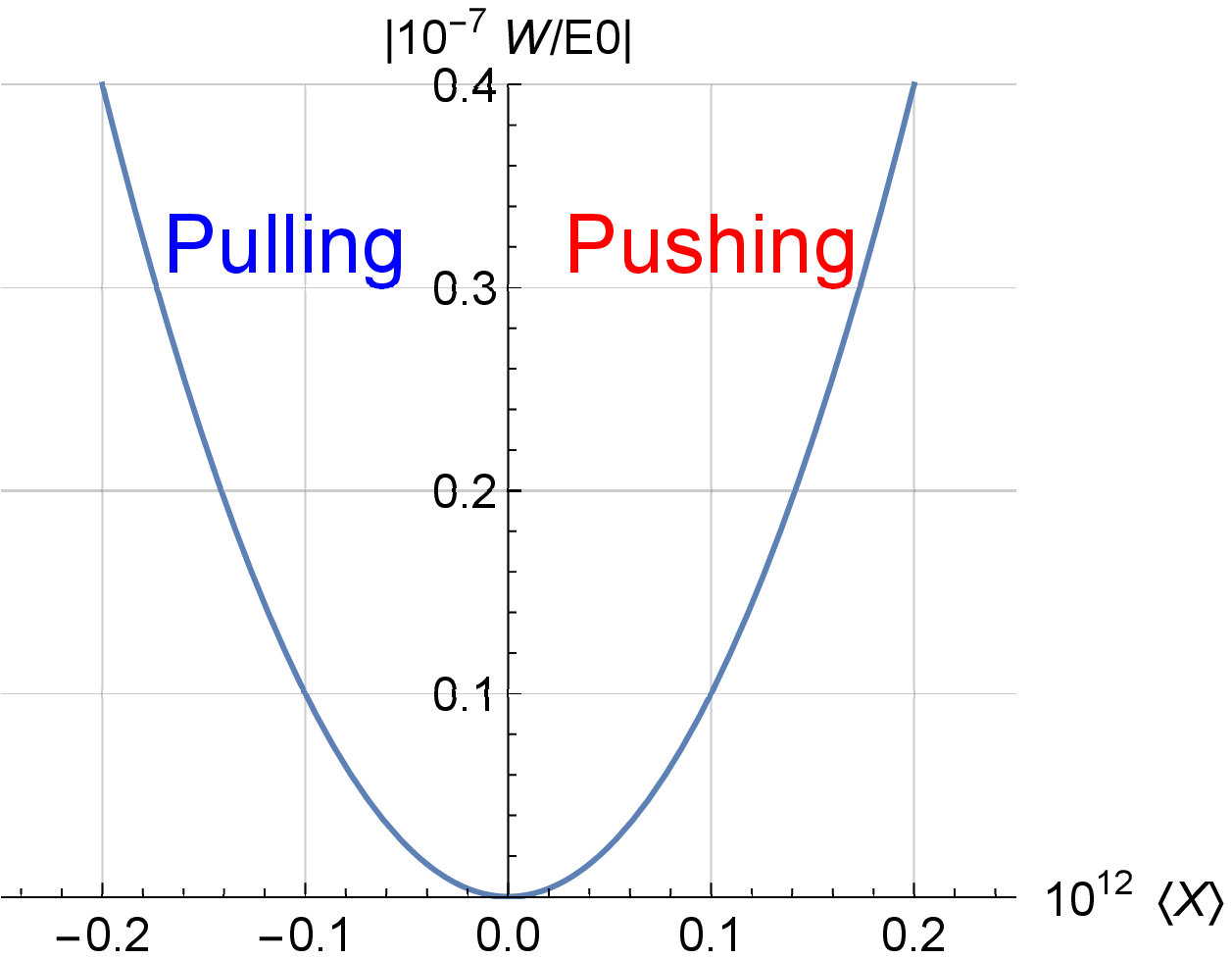}}
\subfigure[]{
\includegraphics[width=0.5\linewidth]{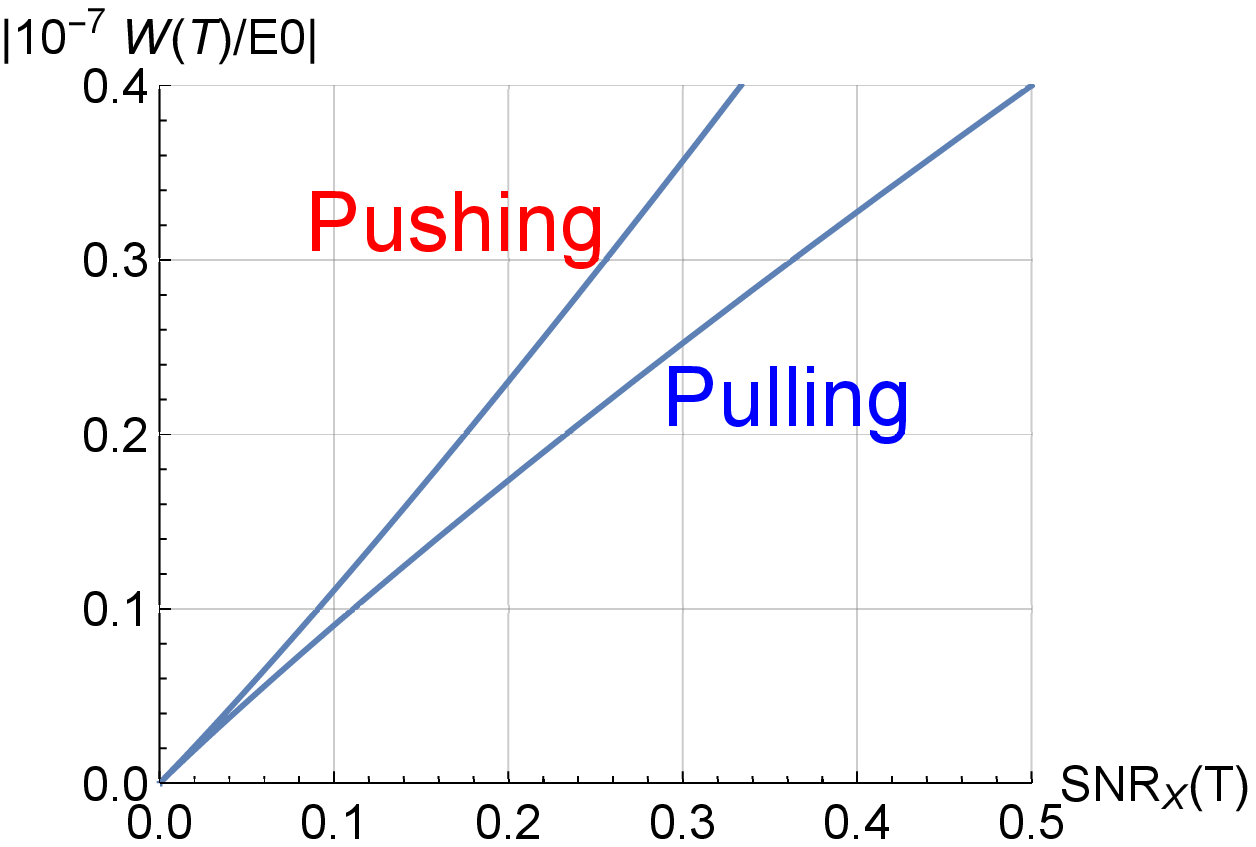}}
\caption{The connection of work, Eq.~\eqref{eq-equilibrium-work}, with the (a) mean value $\qmean{X}$, Eq.~\eqref{eq-equilibrium-variance}, and (b) SNR$_X(T)$, Eq.~\eqref{eq-equilibrium-snr}. The joint parametric dependence of the plotted variables on $T$ is used in the range $T\in (80,120)$ K. Pushing ($T>T_0$) and pulling ($T<T_0$) branches of the dependences are laballed in correspondence to Fig.~\ref{figure-snr}. Energy unit $E_0$ is the energy of the membrane ground state. We used the model \eqref{eq-equilibrium-fT-model1} and the following values \cite{steeneken}, $\alpha \approx 10^{-12}\;{\rm m/K},\; m\omega^2\approx 10^{4}\; {\rm kg/s^2},\;\kappa/m\omega^2\approx 10^{-2},\;T_0\approx 10^{2}\;{\rm K}$.   } \label{figure-work}
\end{figure}

With the change of the bath temperature $T_0\rightarrow T$ the piston drives the membrane into the new average value of the position $\qmean{X}$, Eq.~\eqref{eq-equilibrium-average}. This transition changes as well the sum of the potential energies $E_p^{\rm coh}+E_p^{\rm inc}$, thus the internal energy $U$, Eq.~\eqref{eq-equilibrium-potentials-U}, changes as well. Such a change must be accompanied with the exchange of work or heat with the membrane surroundings according to the first law of thermodynamics \cite{greiner}. The work done by the piston on the membrane during the change $T_0\rightarrow T$ is, cf. Eq.~\eqref{eq-appA-1st-law}
\begin{eqnarray}\nonumber
|W(T_0\rightarrow T)|&=&\frac{f(T)^2}{2m\omega^2}\\
&=&k_BT\frac{{\rm SNR}_X(T)}{2}=|E_p^{\rm coh}(T)|,
\label{eq-equilibrium-work}
\end{eqnarray}
being the change of the coherent part of the potential energy in Eq.~\eqref{eq-equilibrium-coherent-potE}, as anticipated. 
The connection of the work, Eq.~\eqref{eq-equilibrium-work}, with mechanical quantities $\qmean{X}$, Eq.~\eqref{eq-equilibrium-variance}, and SNR$_X(T)$, Eq.~\eqref{eq-equilibrium-snr}, is shown in Fig.~\ref{figure-work}.
This value can be compared to heat $Q(T_0\rightarrow T)=k_B(T-T_0)$, the ``useless'' form of energy, that enters (leaves) the membrane during the bath temperature change $T_0\rightarrow T$. This absorbed (released) heat only changes the $E_p^{\rm inc}(T)$, Eq.~\eqref{eq-equilibrium-incoh-potE}, and the average {\it kinetic} energy (the first term in Eq.~\eqref{eq-equilibrium-average-E}).

\begin{figure}[htb]
\centering \hspace{-0.06\linewidth}
\includegraphics[width=.7\linewidth]{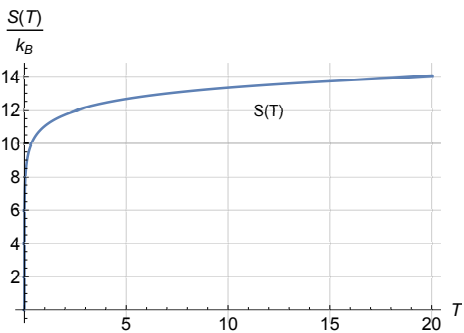}
\caption{The entropy $S$, Eq.~\eqref{eq-equilibrium-potentials-S}, dependence on the thermodynamic temperature $T$ (in Kelvins). We used the model \eqref{eq-equilibrium-fT-model1} and the following values \cite{steeneken}, $\alpha \approx 10^{-12}\;{\rm m/K},\; m\omega^2\approx 10^{4}\; {\rm kg/s^2},\;\kappa/m\omega^2\approx 10^{-2},\;T_0\approx 10^{2}\;{\rm K}$.   } \label{figure-entropy}
\end{figure}
\begin{figure}[htb]
\centering \hspace{-0.06\linewidth}
\includegraphics[width=.7\linewidth]{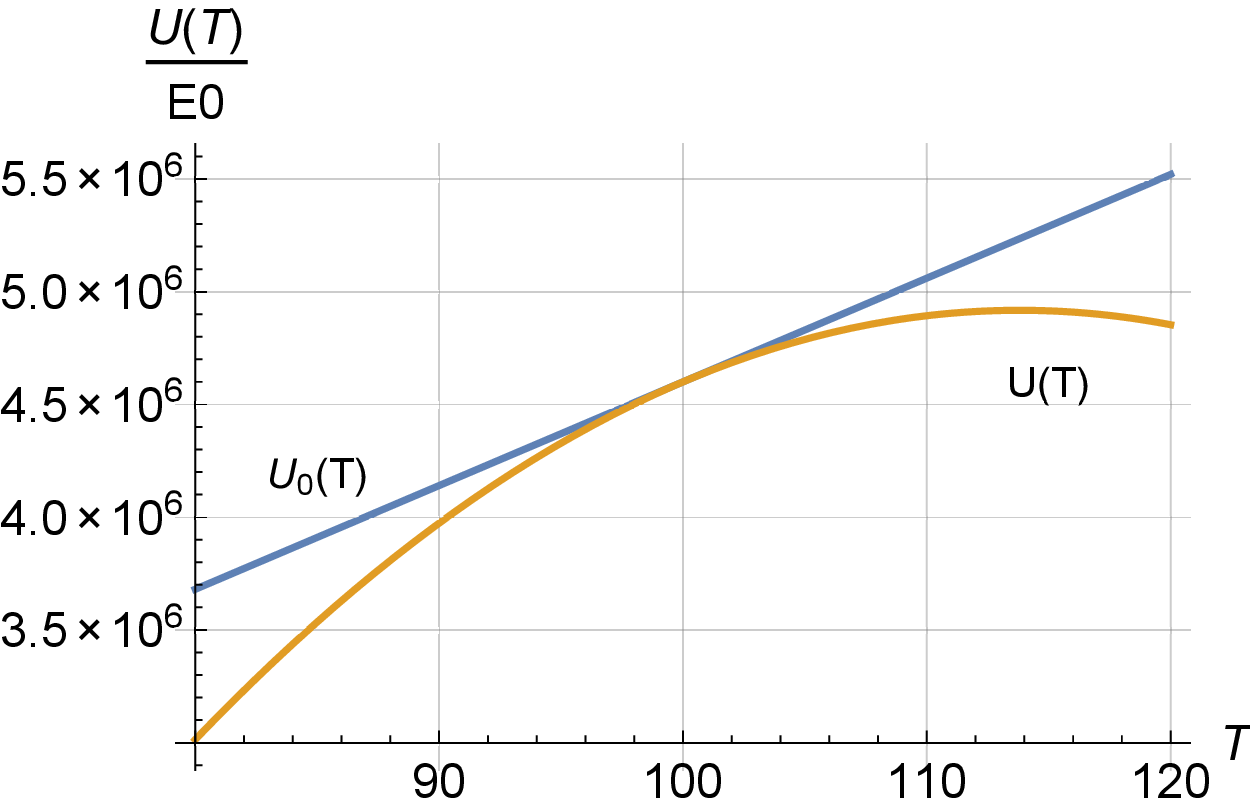}
\caption{The internal energy $U$, Eq.~\eqref{eq-equilibrium-potentials-U},  dependence on the thermodynamic temperature $T$ (in Kelvins). Energy unit $E_0$ is the energy of the membrane ground state. We used the model \eqref{eq-equilibrium-fT-model1} and the following values \cite{steeneken}, $\alpha \approx 10^{-12}\;{\rm m/K},\; m\omega^2\approx 10^{4}\; {\rm kg/s^2},\;\kappa/m\omega^2\approx 10^{-2},\;T_0\approx 10^{2}\;{\rm K}$.   } \label{figure-U}
\end{figure}

For further increase of the SNR$_X(T)$ we can utilize additional cooling mechanism to bring the membrane closer to its ground state. Modeling the cooling mechanism as membrane coupling to a low temperature bath under the quantum optical limit assumption \cite{petrucione}, see Eq.~\eqref{eq-laser-cooled}, we may bring the membrane to the state characterized by the {\it effective}  temperature \cite{aspelmeyer-book} $T^\star \approx\epsilon T,\;\epsilon\ll 1$, Eq.~\eqref{eq-eff-temp-definition}. 
The temperature $T^\star$ is effective because the membrane is not in a thermal equilibrium since it is in contact with two heat baths: the mechanical heat bath with temperature $T$ and the laser ``heat bath'' with temperature $T_L\approx 0$. Nevertheless, the Gaussian form of its stationary state allows for effective parametrization with $T^\star$.
Here, $T$ is still the membrane heat bath temperature. This effective cooling of the membrane thermal fluctuations {\it does not} affect the value of stationary average position, although it {\it decreases} the position variance, see Eq.~\eqref{eq-quantum-stationary}. For results stemming from the stationary membrane state see Appendix~\eqref{app-B}. In this setting the improved SNR$_X^\star$ scales as ${\rm SNR}_X^\star ={\rm SNR}_X/\epsilon$, while the work $W$ done by the piston on the membrane, Eq.~\eqref{eq-equilibrium-work}, remains the same. This effective approach can be used to experimentally verify the low temperature situation, still, however, far from quantum limit. The effective temperature $T^\star$ will change to temperature $T$ only if the temperature of mechanical bath is simultaneously reduced to $T^\star$. This desirable case, without any thermal heating rate, will establish once again the thermal equilibrium and all the results of this section can be directly applied.    

\section{New perspectives on thermal manipulations in optomechanics}
\label{conclusions}

We have described the impact of thermal driving on the optomechanical membrane from a mechanical, as well as thermodynamic point of view. Mechanically, we recognized that one can prepare a low-noise (in the sense of high SNR) mechanical state of the membrane by piston thermal expansion. 
Thermodynamically, such temperature dependent driving has impact on the ground state energy of the membrane and thus modifies significantly the membrane internal energy and its change with the temperature.

We have pointed out that different methods, measuring different aspects of the membrane state, can in principle give different results in the heat capacity measurement. Such contrast allows to access qualitatively different properties of the membrane thermal state and thus could be of imminent interest for further research on heat capacity measurement methods applicable in optomechanical settings. 

Future research activities will be dedicated to study the properties of simple systems with other temperature dependent parameters, e.g., the angular frequency of the membrane. Next step is an investigation of thermally nonlinear potentials to prepare highly nonclassical quantum states of mechanical systems. This new direction of investigation is very stimulating for a proof-of-principle experimental investigation, possibly with levitating nanospheres \cite{gieseler}, or mechanical toroids \cite{harris}.

\section*{Acknowledgments}
M.K. and R.F. acknowledge the support of the project GB14-36681G of the Czech Science Foundation. A.R. gratefully acknowledges support of the project No.~17-06716S of the Czech Science Foundation.

\appendix
\section{Thermodynamics of systems with a temperature dependent Hamiltonians}
\label{sec-app-A}
Our initial assumption is that the form of the  equilibrium Gibbs state (for given temperature $T$) remains unchanged by the explicit temperature $T$ and external parameters $\{x_i\}$ dependence of the system Hamiltonian, i.e., \cite{greiner}
\begin{eqnarray}\nonumber
\op{\rho}&=&\frac{\exp[-\op{H}(T,\{x_i\})/k_BT]}{Z}\\
&=&\exp\left[\frac{F-\op{H}(T,\{x_i\})}{k_BT}\right],\label{eq-appA-gibbs}\\
F&=&-k_BT\ln Z,\;Z={\rm Tr}\left[\exp\left(-\frac{\op{H}(T,\{x_i\})}{k_BT} \right)\right].\left. \right. 
\label{eq-appA-FZ}
\end{eqnarray}
These definitions reflect only the standard normalization condition of the density matrix. The central quantity of our interest is the von Neumann entropy defined in the standard manner \cite{greiner}
\begin{eqnarray}
S&=&-k_B\langle \ln\op{\rho}\rangle =-k_B{\rm Tr}[\op{\rho}\ln\op{\rho}].
\label{eq-appA-S}
\end{eqnarray}
This common definition, making use of the {\it unchanged} state \eqref{eq-appA-gibbs}, keeps also the standard \cite{greiner} connection between the quantities $F$, $S$, $U$ and $T$, namely
\begin{eqnarray}\nonumber
S&=&-k_B{\rm Tr}\left[\op{\rho}\;\frac{(F-\op{H}(T,\{x_i\}))}{k_BT} \right]\\
&=&\frac{\langle\op{H}(T,\{x_i\})\rangle-F}{T}.
\label{eq-appA-SFT-connection}
\end{eqnarray}
This result allows us to interpret $\qmean{\op{H}(T,\{x_i\})}$ as the internal energy $U$ of the system and $F$ as the Helmholtz free energy of the system. With this interpretation we can write the first law of thermodynamics as ${\rm d}U~\equiv~{\rm d}\qmean{\op{H}(T,\{x_i\})}$
\begin{eqnarray}\label{eq-appA-1st-law}
{\rm d}U&=&T{\rm d}S+\frac{\partial \qmean{\op{H}(T,\{x_i\})}}{\partial T}{\rm d}T\\
\nonumber
&+&\sum_i\frac{\partial \qmean{\op{H}(T,\{x_i\})}}{\partial x_i}{\rm d}x_i,
\end{eqnarray}
where we have explicitly kept the temperature dependent term, whereas $x_i$ are all other external parameters of $\op{H}(T,\{x_i\})$, e.g., the frequency $\omega$ for the case of the oscillator \eqref{eq-equilibrium-ham}. The quantity $S$ is the von Neumann entropy \eqref{eq-appA-S}. 

The $T{\rm d}S$ term in \eqref{eq-appA-1st-law} is thus the heat term describing the change of the energy levels populations $p_n(T)$ while energies $E_n(T,\{x_i\})$ are fixed because
\begin{eqnarray}
T{\rm d}S=\sum_n E_n(T,\{x_i\}){\rm d}p_n,
\label{eq-appA-heat}
\end{eqnarray}
where $E_n(T,\{x_i\})$ are the eigenvalues of the Hamiltonian $\op{H}(T)$. 

The last term in Eq.~\eqref{eq-appA-1st-law} represents the work of the external forces on the system. 

The middle term interpretation in Eq.~\eqref{eq-appA-1st-law} can be either as the heat term \cite{shental} or as the work term, as adopt. The importance of this term appears if we Legendre-transform the internal energy $U$ into the free energy $F$ \cite{greiner}
\begin{eqnarray}
\label{eq-appA-dif-F}
{\rm d}F&=&{\rm d}(U-TS)\\\nonumber
&=&-S{\rm d}T+\frac{\partial \qmean{\op{H}(T,\{x_i\})}}{\partial T}{\rm d}T+\sum_i\frac{\partial \qmean{\op{H}(T,\{x_i\})}}{\partial x_i}{\rm d}x_i.
\end{eqnarray}
Form of Eq.~\eqref{eq-appA-dif-F} suggests that the von Neumann entropy $S$ is now given as, cf. Eq.~\eqref{eq-equilibrium-potentials-S},
\begin{eqnarray}
S=-\frac{\partial [F-\qmean{\op{H}(T,\{x_i\})}]}{\partial T}.
\end{eqnarray}
\section{Laser cooled membrane}
\label{app-B}
In this appendix, we describe the situation when the membrane is cooled closer to its ground state by means of the thermal energy dissipation. 
The Langevin-type equations of motion for the operators describing the underdamped membrane driven by the thermal piston are, with the substitution $\op{X}_1=\op{X}-f(T)/(m\omega^2)$,
\begin{eqnarray}
\nonumber
\dot{\op{X}}_1&=&-\frac{\gamma_L}{2}\op{X}_1+\frac{\op{P}}{m}+\sqrt{\frac{\hbar}{2m\omega}}(\op{L}^\dagger +\op{L}) \\
\nonumber
\dot{\op{P}}&=&-m\omega^2 \op{X}_1-\left(\frac{\gamma_L}{2}+\frac{\gamma_M}{m} \right)\op{P} \\
&&+\sqrt{2\gamma_M k_BT}\xi_M +i\sqrt{\frac{\hbar m\omega}{2}}(\op{L}^\dagger -\op{L}),
\label{eq-laser-cooled}
\end{eqnarray}
where $\gamma_M$ is the mechanical damping constant characterizing the Brownian motion of the membrane,
$\gamma_L$ is the effective damping constant due to the laser cooling and $\op{L}$ the Langevin force operator due to the laser cooling. We will omit the ``caret" for the operators in the following. These equations are valid under assumptions $\gamma_L\ll\omega$ (quantum optical limit) and $\gamma_M\ll m\omega$ (underdamped Brownian motion). 
Such dynamical equations lead the membrane towards the Gaussian stationary state, $\rho_{{\rm st}}$, characterized fully by the first and second moments
\begin{eqnarray}
\label{eq-quantum-stationary-average}\langle X_1\rangle &=&\langle P\rangle=0,\\
\langle X^2_1\rangle &=&\frac{1}{m^2\omega^2}\langle P^2\rangle
=\frac{\hbar}{2m\omega}\left(1-\epsilon \right)+\epsilon\frac{k_BT}{m\omega^2},\label{eq-quantum-stationary}\\
\epsilon &=&\frac{\gamma_M}{m\gamma_L}\ll 1,
\end{eqnarray}
where we have assumed $\gamma_M/m\ll \gamma_L\ll\omega$, consistently with the derivation assumptions of Eq.~\eqref{eq-laser-cooled}. 

Equation~\eqref{eq-quantum-stationary} allows for the definition of the effective temperature $T^\star$ characterizing fully the membrane state (effectively of the canonical form). 
The effective temperature can be introduced through the result valid for quantum harmonic oscillator
\begin{eqnarray}
\langle X^2_1\rangle &=&\frac{1}{m^2\omega^2}\langle P^2\rangle \equiv\frac{\hbar}{2m\omega}\left(2n^\star+1\right),\\
n^\star &\equiv &\left[\exp\left(\frac{\hbar\omega}{k_BT^\star} \right)-1\right]^{-1}=\epsilon\frac{k_BT}{\hbar\omega},\;n^\star\gg 1,\\
\label{eq-eff-temp-definition} T^\star &=& \frac{\hbar\omega}{k_B}\left[\ln\left(\frac{1}{n^\star}+1 \right) \right]^{-1}\approx \frac{\hbar\omega}{k_B}n^\star\\\nonumber
&\approx & \epsilon T,\;n^\star\gg 1.
\end{eqnarray}

The canonical form of the Gaussian state with parameters as in Eqs.~\eqref{eq-quantum-stationary-average}, \eqref{eq-quantum-stationary} allow for formal definition of certain (originally equilibrium) thermodynamic quantities. From these, we choose state functions, namely the internal energy $U=\langle H(T)\rangle$, entropy $S=-k_B\qmean{\ln \rho_{{\rm st}}}$, and free energy $F=U-T^\star S$, where the subscript ``st'' stands again for stationary, 
\begin{eqnarray}
\label{eq-noneq-quantities-U}
U&=&k_BT^\star-\frac{f(T)^2}{2m\omega^2},\\
S&=&k_B\left[1-\ln\left(\frac{\hbar\omega}{k_BT^\star} \right) \right],
\\
\label{eq-noneq-quantities}
F&=&k_BT^\star\ln\left(\frac{\hbar\omega}{k_BT^\star} \right)-\frac{f(T)^2}{2m\omega^2}.
\end{eqnarray}

Partial derivatives of the quantities \eqref{eq-noneq-quantities-U}--\eqref{eq-noneq-quantities} with respect to the $T$ have the same dependence as in the equilibrium case, but not the same meaning. For example, cf. Eq.~\eqref{eq-equilibrium-heat-cap}, 
\begin{eqnarray}
\frac{{\rm d}U}{{\rm d}T} =\epsilon k_B -\frac{1}{2m\omega^2}\frac{{\rm d}f^2}{{\rm d}T},
\label{eq-noneq-capacity-U}
\end{eqnarray}
is not the constant-frequency heat capacity. It is more a coefficient describing the membrane ability to capture a part of the heat flow, thus increasing its internal energy $U$.

\end{document}